\documentclass[twocolumn,english,aps]{revtex4}
\usepackage[T1]{fontenc}
\usepackage[latin1]{inputenc}
\usepackage{graphicx}
\usepackage{amssymb}

\makeatletter

\@ifundefined{definecolor}
 {\@ifundefined{definecolor}
 {\usepackage{color}}{}
}{}

\makeatletter

\makeatletter

\usepackage{soul}

\makeatother

\makeatother

\makeatother

\usepackage{babel}

\makeatother

\usepackage{babel}

\begin{document}

\title{Compressive ghost imaging}

\author{Ori Katz}

\email{ori.katz@weizmann.ac.il}

\affiliation{Department of Physics of Complex Systems, The Weizmann Institute
of Science, Rehovot, Israel}

\author{Yaron Bromberg}

\affiliation{Department of Physics of Complex Systems, The Weizmann Institute
of Science, Rehovot, Israel}

\author{Yaron Silberberg}

\affiliation{Department of Physics of Complex Systems, The Weizmann Institute
of Science, Rehovot, Israel}
\begin{abstract}
We describe an advanced image reconstruction algorithm for
pseudothermal ghost imaging, reducing the number of measurements
required for image recovery by an order of magnitude. The algorithm
is based on \textit{compressed sensing}, a technique that enables
the reconstruction of an $N$-pixel image from much less than $N$
measurements. We demonstrate the algorithm using experimental data
from a pseudothermal ghost-imaging setup. The algorithm can be
applied to data taken from past pseudothermal ghost-imaging
experiments, improving the reconstruction's quality.

\end{abstract}
\maketitle Ghost imaging (GI) has emerged a decade ago as an imaging
technique which exploits the quantum nature of light, and has been
in the focus of many studies since \cite[and references
therin]{GattiReview}. In GI an object is imaged even though the
light which illuminates it is collected by a single-pixel detector
which has no spatial resolution (a bucket detector). This is done by
correlating the intensities measured by the bucket detector with an
image of the field which impinges upon the object. GI was originally
performed using entangled photon pairs \cite{PitmanPRA95}, and later
on was realized with classical light sources
\cite{Boyd02,XrayPRL,GattiPRL05,ShihPRL05}. The demonstrations of GI
with classical light sources, and especially pseudothermal sources,
triggered an ongoing effort to implement GI for various sensing
applications \cite{XrayPRL,ShihSoldierPRA08}. However, one of the
main drawbacks of pseudothermal GI is the long acquisition times
required for reconstructing images with a good signal-to-noise ratio
(SNR) \cite{GattiReview,ShapiroSNR}.

In this work we propose an advanced reconstruction algorithm for
pseudothermal GI, which reduces significantly the required
acquisition times. The algorithm is based on \textit{compressed
sensing} (or \textit{compressive sampling}, CS)
\cite{CSintro,Imaging via CS}, an advanced sampling and
reconstruction technique which has been recently implemented in
several fields of imaging. Examples for such are magnetic resonance
imaging \cite{MRI CS}, astronomy \cite{CS in Astronomy}, THz imaging
\cite{THz CS}, and single-pixel cameras \cite{single pixel camera}.
The main idea behind CS is to exploit the redundancy in the
structure of most natural signals/objects to reduce the number of
measurements required for faithful reconstruction. Here we show that
applying a CS-based reconstruction algorithm to data taken from
conventional pseudothermal GI measurements dramatically improves the
SNR of the reconstructed images and thus allows for shorter
acquisition times.

\begin{figure}
\centering{}\includegraphics[width=0.9\columnwidth]{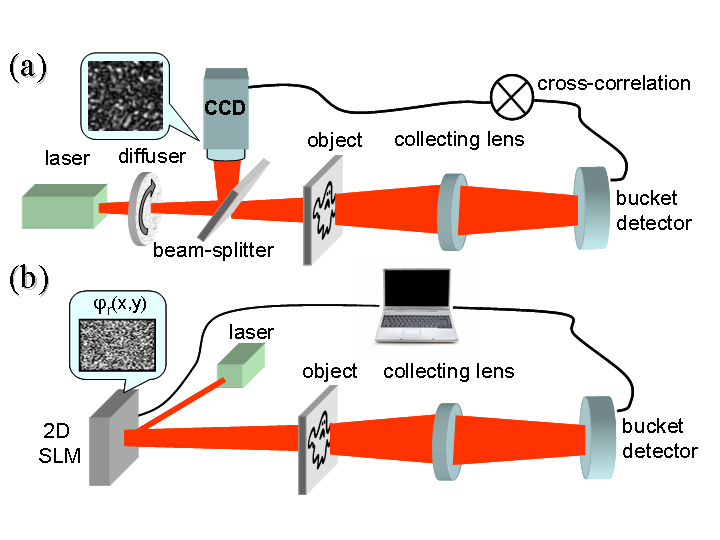}
\caption{\label{fig:setup} (Color online) (a) Standard pseudothermal
GI two-detectors setup. A copy of the speckle field which impinges
on the object is imaged with a CCD camera, and correlated with the
intensity measured by a bucket detector. (b) The computational GI
single-detector setup used in this work. A pseudothermal light beam
is generated by applying controllable phase masks $\varphi_{r}(x,y)$
with a spatial light modulator (SLM). The object image is obtained
by correlating the intensity measured by the bucket detector, with
the \emph{calculated} field at the object plane. }
\end{figure}

In conventional pseudothermal GI, an object is illuminated by a
speckle field generated by passing a laser beam through a rotating
diffuser [Fig.~1(a)]. For each phase realization $r$ of the
diffuser, the speckle field $I_{r}(x,y)$ which impinges on the
object is imaged. This is done by splitting the beam before the
object to an 'object arm' and a 'reference arm', and placing a CCD
camera at the reference arm. At the object arm, a bucket detector
measures the total intensity $B_{r}$ which is transmitted through
the object, described by a transmission function
$T(x,y)$:\begin{equation} B_{r}=\int
dxdyI_{r}(x,y)T(x,y).\label{eq:Br}\end{equation}. To reconstruct the
object's transmission function, the bucket detector measurements are
cross-correlated with the intensities measured at the reference
arm:\begin{eqnarray} T_{GI}(x,y) & = &
\frac{1}{M}\sum_{r=1}^{M}(B_{r}-\langle
B\rangle)I_{r}(x,y),\label{eq:T_GI}\end{eqnarray}

\noindent where $\langle\cdot\rangle\equiv\frac{1}{M}\sum_{r}\cdot$
denotes an ensemble average over $M$ realizations (measurements).
From Eq.~(\ref{eq:T_GI}) one can see that the image is obtained by a
linear superposition of the intensity patterns $I_{r}(x,y)$ with the
appropriate weights $B_{r}-\langle B\rangle$. Each bucket
measurement $B_{r}$ is the overlap between the object and the
illumination pattern [Eq.~(\ref{eq:Br})]. Thus, the GI measurement
process is in essence a vector projection of the object transmission
function $T(x,y)$ over $M$ different random vectors $I_{r}(x,y)$.

The GI linear reconstruction process has no assumptions on the
to-be-resolved object. Thus if the number of resolution cells
(speckles) which cover the object is $N$, one needs at least $M=N$
different intensity patterns in order to reconstruct the object (the
measurement's Nyquist limit). In fact, since the different intensity
patterns $I_{r}(x,y)$ overlap, $M\gg N$ measurements are needed to
meet $SNR\gg1$ \cite{ShapiroSNR,GattiReview}. However, any prior
information on the structure of the object could significantly
reduce the number of measurements required for a faithful
reconstruction. Remarkably, for most imaging tasks such information
exists: natural images are sparse, that is, they contain many
coefficients close to or equal to zero when represented in an
appropriate basis (e.g. the discrete cosine transform (DCT)). This
fact is at the core of modern lossy image compression algorithms,
such as JPEG \cite{Imaging via CS}. The main idea behind CS is to
exploit this sparsity/compressibility to reduce the number of
measurements needed for faithful image recovery.

CS reconstruction algorithms search for the most sparse image in the
compressible basis which fulfills the $M<N$ random projections
measured. It requires solving a convex optimization program, seeking
for the image $T_{CS}(x,y)$ which minimizes the $L_{1}$-norm in the
sparse basis (i.e. the sum of the absolute values of the transform
coefficients) \cite{CSintro,Imaging via CS}:\begin{eqnarray}
T_{CS} & = & T'\text{ which minimizes: }||\Psi\left\{ T'(x,y)\right\} ||_{L_{1}}\nonumber \\
\text{subject} & \text{to} &  \int dxdyI_{r}(x,y)T'(x,y)=B_{r}, \;
\forall _{r=1..M}\label{eq:T_CS}\end{eqnarray}

\noindent where $B_r$ are the $M$ gathered projections measurements,
and $\Psi$ is the transform operator to the sparse basis (e.g.
2D-DCT). Finding the image with the minimal $L_{1}$-norm can be
realized as a linear program, for which efficient solution methods
exist. According to CS theory, one can reconstruct compressible
images characterized by $K$ transform coefficients with high
fidelity, using just $M \geq O\left(K log(N/K)\right)$ random
measurements, where $N$ is the total number of resolution cells in
the image. Reconstruction of natural images using CS was
demonstrated using $M\lesssim N/2$ measurements \cite{Imaging via
CS,MRI CS, single pixel camera}. This sub-Nyquist acquisition
results from exploiting the image natural sparsity. We note that
since the vectors $I_{r}(x,y)$ are random, they are most likely
linearly independent, and therefore $M\geq N$
projection-measurements span the image dimensionality. Thus, the
image can be reconstructed (without exploiting the image sparsity)
by solving a set of $M$ linear equations using conventional linear
least-squares methods. Such a linear algebra based reconstruction
outperforms the standard GI reconstruction when $M\geq N$, and gives
a perfect result in the absence of measurement noise.

To experimentally demonstrate CS reconstruction in GI, we have used
the computational GI setup presented in \cite{BrombergCGI}
[Fig.~1(b)]. Computational GI is a variant of the standard
two-detectors pseudothermal GI, where the rotating diffuser is
replaced by a computer controlled spatial light modulator (SLM)
\cite{shapiroCGI}. Knowing the applied SLM phase mask for each
realization $\varphi_{r}(x,y)$, the intensity of the field at the
reference arm $I_{r}(x,y)$ is computed using the Fresnel-Huygens
propagator, instead of it being measured as in conventional GI. It
is important to note that CS reconstruction can be applied to any
form of pseudothermal GI. It makes no difference if the reference
intensity patterns are computed or measured.

\begin{figure}
\centering{}\includegraphics[width=0.99\columnwidth]{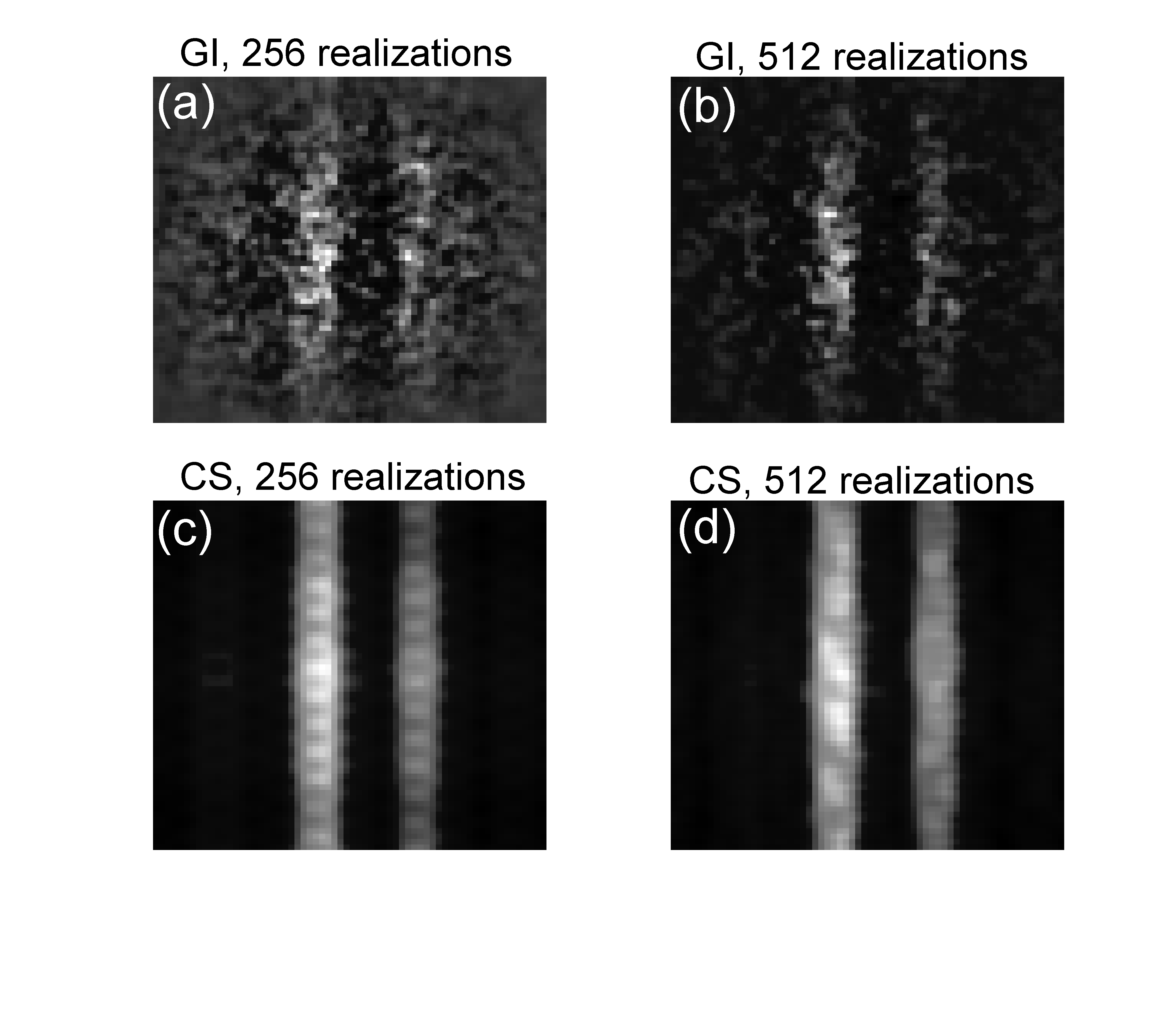}
\caption{\label{fig: Doubleslit} Experimental reconstruction of a
double-slit transmission plate. Top panel: conventional GI
reconstruction with 256 realizations (a), and 512 realizations (b);
Bottom panel (c,d): CS reconstruction using the same experimental
data as in (a) and (b). The increase in SNR using CS reconstruction
is by a factor of $\sim\times4$ in both cases.}
\end{figure}

The reconstruction results for a double slit transmission plate
(width $220\mu m$, separation $500\mu m$), using $M=256$ and $M=512$
realizations are summarized in Fig.~2. The results of conventional
GI reconstruction are plotted in Fig.~2(a-b), and the CS
reconstructions \textit{using the same set of measured data} are
plotted in Fig.~2(c-d). To quantify the improvement gained by
utilizing CS reconstruction, we have calculated the mean SNR of the
reconstructed images. The signal was taken as the difference between
the mean intensity of the bright slit and the dark background, and
the noise was taken as the standard deviation of the dark background
pixels. The calculated SNR for the CS reconstruction using 256
realizations is  $\times4.4$ times higher than for the standard GI
reconstruction, and is $\times4$ times higher in the 512
realizations case. Since the SNR in conventional GI scales as the
square-root of the number of realizations
\cite{ShapiroSNR,GattiReview}, our results imply that CS allows for
an order of magnitude faster image acquisition, making it attractive
for practical imaging tasks. The reconstructions fidelity was
estimated by calculating the mean-square error (MSE) of the
reconstructions compared to a reference image $T_{ref}$, measured
directly by a transmission microscope. The MSE given by
$\frac{1}{N_{pix}}\sum_{i,j}
(T_{CS/GI}(x_i,y_j)-T_{ref}(x_i,y_j))^2$ is $0.12$ for GI and $0.04$
for CS using $512$ realizations, and $0.14$ for GI and $0.05$ for CS
using $256$ realizations. The summation is done over all the image
pixels $N_{pix}$.

The pixel-resolution of the calculated speckle-field image
$I_{r}(x,y)$ used for the reconstructions was $64\times64$ pixels
($N_{pix}=4096$). At this resolution the speckles full-width at
half-max (FWHM) was $1.53$ pixels, yielding $N=1750$ resolution
cells covering the object (the measurement's Nyquist limit).
Therefore in Fig.~2(a,c) the number of measurements used for the
reconstructions is $15\%$ of the Nyquist limit, and is $30\%$ of the
Nyquist limit in Fig.~2(b,d). The pixel-resolution was chosen such
that the individual speckles are resolved (pixel size < speckle
size), yet the required computational resources are minimized. For
the CS reconstruction we have utilized the gradient projection for
sparse reconstruction (GPSR) algorithm \cite{GPSR algorithm},
minimizing the $L_{1}$-norm in the 2D-DCT domain. This algorithm
follows Eq. (\ref{eq:T_CS}), but considers the presence of noise in
each measurement $B_{r}$, by relaxing the equality constraint.

To verify the applicability of CS reconstruction for more general
images, we have imaged a transmission plate of the Hebrew letter
Aleph ($\aleph$). The results for both GI and CS reconstructions
using the same set of 1024 measurements are presented in
Fig.~3(a,c). The reference data size used in this case was
$70\times76$ pixels, and the speckles size was 2.01 pixels FWHM
($N=1330$). The calculated SNR for the CS reconstruction was
$\times3.5$ times higher than the GI SNR, and the calculated MSE was
$0.05$ and $0.1$, respectively. Finally, we demonstrate sub-Nyquist
CS-GI reconstruction of a natural grayscale image, by reconstructing
a $76\times70$ pixel image, containing $N=1330$ resolution cells
from $800$ simulated measurements, obtained by multiplying the
speckle patterns used in the Aleph reconstruction experiment by the
grayscale image values. The obtained MSE is $0.09$ for the GI
reconstruction and $0.005$ for the CS reconstruction [Fig.~3(b,d)].

\begin{figure}
\centering{}\includegraphics[width=0.99\columnwidth]{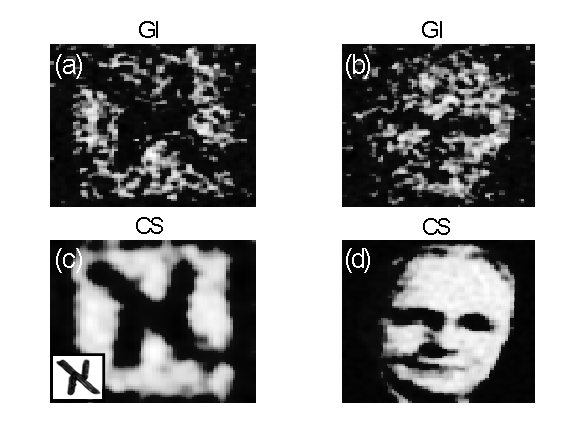}
\caption{\label{fig: Aleph} (a) Experimental GI reconstruction of a
transmission plate of the Hebrew letter Aleph ($\aleph$) from 1024
measurements. (c) same as (a) but utilizing CS reconstruction,
yielding $\times3.5$ times higher SNR. (inset: the object's
transmission image). (b,d) Simulated GI and CS reconstructions of a
$70\times76$ pixels grayscale portrait of H.Nyquist, using $800$
measurements (60\% the Nyquist limit).}
\end{figure}

In conclusion, we have shown that by employing notions from CS
theory in a GI reconstruction algorithm, one can boost the recovered
image quality. CS unleashes the full potential of the random
projections measurement process of pseudothermal GI. It enables
image reconstruction with far less measurements than is possible
with conventional GI, and in some scenarios, with a scanning beam
imaging setup. CS therefore holds potential for future
implementations of GI in practical applications such as LIDAR. The
presented algorithm can be applied to any pseudothermal GI data
taken in the past, yielding superior reconstruction. Moreover, since
computational GI allows for scanning-less three-dimensional (3D)
image reconstruction \cite{BrombergCGI}, one may consider applying
CS-GI to reconstruct 3D objects utilizing sparsity in the 3D-DCT
domain or any other 3D-sparse transform basis.

\newpage
\begin{acknowledgements}

We thank Igor Carron, Justin Romberg, Amnon Amir and Dror Baron for
helpful discussions, and KFC and Wim for the arXiv blog thread.
\end{acknowledgements}

\end{document}